# An Automatic Pipeline for the Integration of Python-Based Tools into the Galaxy Platform: Application to the *anvi'o* Framework

Fabio Cumbo[1] ([0000-0003-2920-5838](#)), Jayadev Joshi[1] ([0000-0001-7589-5230](#)), and Daniel Blankenberg[1,2,*] ([0000-0002-6833-9049](#))

[1] Center for Computational Life Sciences, Cleveland Clinic Research, Cleveland Clinic, Cleveland, OH 44195, USA
[2] Department of Molecular Medicine, Cleveland Clinic Lerner College of Medicine, Case Western Reserve University, Cleveland, OH 44195, USA

* To whom correspondence should be addressed:
Daniel Blankenberg[1,2], Center for Computational Life Sciences, Cleveland Clinic Research, Cleveland Clinic, 9500 Euclid Avenue, NA2, Cleveland, OH 44195, USA.
Email: [blanked2@ccf.org](mailto:blanked2@ccf.org)

## ABSTRACT

**Background** – The integration of command-line tools into the Galaxy platform is crucial for making complex computational methods accessible to a broader audience and ensuring reproducible research. However, the manual development of tool wrappers (i.e., the XML files that define the user interface and execution logic in Galaxy) is a time-consuming, error-prone, and knowledge-intensive process. This bottleneck significantly affects the rapid deployment of new and updated tools, creating a gap between tool development and its availability to the scientific community.

**Method** – We have developed a novel, automated approach that directly translates Python tool interfaces into Galaxy-compliant tool wrappers. Our method leverages the *argparse* library, a standard for command-line argument parsing in Python. By embedding structured metadata within the *metavar* attribute of input and output arguments, our system programmatically parses the tool's interface to extract all necessary information. This includes parameter types, data formats, help text, and input/output definitions. The system then uses this information to automatically generate a complete and valid Galaxy tool XML wrapper, requiring no manual intervention.

**Results** – To validate the scalability and effectiveness of our approach, we applied it to the *anvi'o* framework, a comprehensive and complex bioinformatics platform comprising hundreds of individual programs. Our method successfully parsed the *argparse* definitions for the entire *anvi'o* suite and generated functional Galaxy tool wrappers. The resulting integration allows for the seamless execution of *anvi'o* workflows within the Galaxy environment, demonstrating that our system can handle large-scale use cases efficiently and accurately.

**Conclusions** – This work presents a significant advancement in the automation of tool integration for scientific workflow systems. By establishing a convention-based approach using Python's *argparse* library, we have created a scalable and generalizable solution that dramatically reduces the effort required to make command-line tools available in Galaxy.

## INTRODUCTION

The advent of powerful computational methods and high-throughput data generation has fundamentally transformed nearly every scientific field, leading to what is known as the era of big data [1–5]. Disciplines ranging from the life sciences [6] and computational chemistry [7] to machine learning [8] and social sciences [9] now generate vast and complex datasets at an unprecedented rate. The primary challenge has shifted from data generation to data analysis and interpretation, making computational science an indispensable discipline for modern research across numerous domains [10]. To navigate this landscape, scientists rely on a diverse ecosystem of specialized command-line tools, each designed to perform specific tasks, from data preprocessing [11] and statistical modeling [12] to complex simulations [13].

While powerful, the command-line interface of these tools often presents a significant barrier to entry for many domain experts and researchers who may lack extensive computational training. Stringing together multiple tools into a coherent analysis pipeline requires scripting knowledge, meticulous record-keeping, and a deep understanding of software dependencies. These complexities contribute to the "reproducibility crisis" in science, where the results of many studies are difficult or impossible to reproduce independently [14–16]. This challenge underscores the critical need for platforms that not only simplify complex analyses but also enforce and document every step of the process to ensure transparency and robustness.

To address these challenges, scientific workflow management systems like Galaxy [17] have become cornerstones of computational research. These platforms provide graphical user interfaces that enable researchers to build, execute, and share complex, multi-step analyses in a reproducible manner. The Galaxy platform, in particular, has distinguished itself through its strong emphasis on accessibility and its active global community. As a web-based system, Galaxy allows researchers to perform sophisticated data analysis without writing code, managing software installations, or provisioning high-performance computing resources [18]. Its intuitive interface and extensive library of integrated tools have made it a vital resource for tens of thousands of scientists worldwide [19]. Furthermore, Galaxy's architecture is intrinsically aligned with the FAIR (Findable, Accessible, Interoperable, and Reusable) data principles [20], providing a framework where data, tools, and workflows can be easily shared, inspected, and re-executed, thus promoting a more open and collaborative scientific culture.

However, the utility of the Galaxy ecosystem is fundamentally dependent on the availability of tools within it. The process of integrating a new command-line tool into Galaxy (a process known as wrapping) remains a significant and persistent challenge. This integration requires the manual creation of a specific XML file that serves as a bridge between the tool's command-line interface and the Galaxy user interface. Crafting this wrapper is a non-trivial task. It is a time-consuming, error-prone, and knowledge-intensive process that requires expertise in both the underlying tool and the specific syntax of Galaxy's XML schema. This bottleneck creates a substantial lag between the development of novel computational methods and their availability to the broader scientific community that relies on platforms like Galaxy. Consequently, cutting-edge tools can remain inaccessible for months or even years, affecting the pace of scientific discovery.

To quantify the scale of this challenge within the Python ecosystem, we conducted a programmatic analysis of the main Galaxy ToolShed [21]. A tool was classified as Python-based using a set of heuristics, including the presence of *python* in its *<requirements>* tags, dependencies on common Python packages managed by Conda [22] or PyPI, the Python Package Index, or the direct invocation of the Python interpreter within the *<command>* block of the tool wrapper. Our analysis of over 10,000 tool wrappers revealed that the vast majority are dependent on the Python ecosystem. This finding underscores the central role that Python plays in modern bioinformatics and highlights the immense potential impact of a dedicated, automated solution. The sheer volume of Python-based tools in the ToolShed represents a substantial opportunity for automation, validating the need for a scalable integration solution specifically targeting this ecosystem.

To overcome this critical bottleneck, automated solutions are essential. Here, we introduce a novel, scalable, and domain-agnostic framework for the automated generation of Galaxy tool wrappers directly from the source code of Python-based tools. Our method establishes a simple yet powerful convention that leverages the standard argparse library [23], a ubiquitous component for argument parsing in modern Python applications. By embedding structural metadata within the metavar attribute of command-line arguments, our system can programmatically parse a tool's complete interface, including its parameters, data types, and help text, to automatically generate a fully-featured and valid Galaxy tool wrapper.

To demonstrate the power and scalability of our approach, we applied it to the anvi'o framework [24,25], a comprehensive, state-of-the-art bioinformatics platform for the analysis of 'omics data. Comprising hundreds of distinct yet interconnected programs, anvi'o represents a formidable integration challenge due to its complex data structures, numerous inter-tool dependencies, and inclusion of interactive visualization capabilities. It thus serves as an ideal case study to prove the robustness of our domain-agnostic method. Our framework successfully generated functional Galaxy wrappers for the entire anvi'o suite, effectively bridging the gap between this powerful command-line environment and the accessible Galaxy ecosystem. This work not only provides a practical solution to a long-standing problem but also proposes a standardized method that can dramatically accelerate tool integration across all scientific fields, fostering a more dynamic and responsible scientific software ecosystem. This paper will detail the architecture of our framework, validate its performance on the anvi'o suite through several case studies of increasing complexity, and discuss the broader implications for the scientific software community.

**Related Work**

The challenge of integrating command-line tools into scientific workflow management systems is a well-recognized problem, and several strategies have been developed to address it. A prominent approach is the adoption of standardized tool description formats, such as the Common Workflow Language (CWL) [26], which provides a powerful, declarative syntax for describing how a tool should be executed, including its inputs, outputs, and runtime requirements. While this enables portability across different workflow platforms that support the standard, it requires developers to create and maintain a separate, often verbose, CWL definition file for each tool. Other popular workflow systems, like Nextflow [27] and Snakemake [28], address tool integration through their respective domain-specific languages, which offer flexibility but require developers to write custom scripts to define the execution logic for each component in a pipeline.

Within the Galaxy ecosystem, the *planemo* suite of tools [29] has become the standard for assisting in the manual development of tool wrappers. It provides essential utilities for creating wrapper boilerplate, linting XML files to ensure they adhere to best practices, and developing test cases. While indispensable for high-quality tool development, *planemo* is designed to support and validate the manual wrapping process rather than automate the initial generation of the wrapper from a tool's source code.

Our framework presents a distinctive and complementary approach. Instead of relying on external definition files like CWL or custom scripting, our method programmatically introspects the tool's native command-line interface via its *argparse* definitions. By establishing a lightweight, non-intrusive convention using the *metavar* attribute, we enable the direct translation of a tool's interface into a complete Galaxy wrapper. This approach is specifically designed to be developer-friendly for the vast Python ecosystem, lowering the barrier to entry by embedding the necessary metadata within the existing code structure, thereby automating the most time-consuming step of the integration process.

**MATERIALS AND METHODS**

The primary objective of this work is to address the significant bottleneck in scientific software deployment by creating a generalizable and automated pipeline for converting Python-based command-line tools into fully-featured Galaxy tool wrappers. Our framework is designed to be domain-agnostic, capable of processing any Python tool that utilizes the standard *argparse* library for its command-line interface. The core strategy relies on programmatic introspection of the target tool's source code. By dynamically executing the script in a controlled environment, our system intercepts the *argparse* definitions to build a structured representation of the tool's interface. This process is guided by a simple but powerful convention: the use of the *metavar* attribute within an argument's definition to declare its semantic data type. This structured information is then passed to a templating engine, which generates a complete and valid Galaxy tool XML file. The system is architecturally composed of three primary components, which will be detailed in subsequent sections: (i) a dynamic Python script introspector responsible for capturing argument definitions, (ii) a parameter classification engine that maps these arguments to specific Galaxy parameter types based on *metavar* values and argument names, and (iii) a Galaxy XML generator that renders the final wrapper using Jinja2 templates [30].

**Introspection of Python Tool Interfaces**

A central challenge in automated wrapper generation is extracting the complete command-line interface from a tool without requiring manual inspection or modification of its source code. Our framework achieves this through a process of dynamic script introspection. For each target Python script, the framework first reads the entire file into memory. It then performs a series of targeted, in-memory string replacements to prepare the code for execution in a controlled environment. Crucially, all references to the standard *argparse.ArgumentParser* are replaced with a custom-built *FakeArg* class. The main execution block (i.e., "*if __name__ == '__main__':*") is redefined as a function, and the final call that would normally parse arguments and run the tool (i.e., *parser.parse_args()*) is replaced with a return statement that passes the parser object out. This modified code is then executed using Python's *exec* function, which allows our framework to instantiate the parser and capture all of its argument definitions without running the tool's primary logic.

The core of this introspection mechanism is the *FakeArg* class, which is designed to act as a stand-in for the native *argparse.ArgumentParser*. It mirrors the essential methods of the original class but implements a different internal behavior. Instead of preparing to parse command-line arguments, the *FakeArg.add_argument()* method simply records all of the properties passed to it, including the argument's flags (e.g., *-o*, *--output*, etc.), its *help* text, *default* value, *required* status, and *metavar*, into an internal list. This effectively transforms the argument definition phase of the script into a data-gathering process. By the end of the controlled execution, the *FakeArg* object contains a complete and structured representation of the tool's entire command-line interface, which serves as the raw input for the subsequent classification and XML generation stages.

To handle more complex *argparse* features, such as argument groups, the framework employs an additional layer of code manipulation. The system detects when an argument group is created (e.g., *group = parser.add_argument_group(...)*) and identifies the variable name assigned to it. It then replaces all subsequent calls to *add_argument* on that group object (e.g., *group.add_argument(...)*) with an equivalent call on the main parser object. This strategy effectively flattens the argument structure, ensuring that all defined parameters are collected in a single, unified list within the *FakeArg* object, which greatly simplifies the downstream processing logic.

**Parameter Classification and Mapping**

Once the tool's interface has been captured, the framework must translate each command-line argument into its corresponding Galaxy parameter type. This is achieved through a multi-tiered classification engine that prioritizes semantic information when available. The central convention of our framework is the use of the *metavar* attribute in an *argparse* definition to explicitly declare the argument's type. We established a primary mapping, stored in a Python dictionary, that links specific *metavar* strings to specialized parameter-handling classes. For example, an argument defined with *metavar='PROFILE_DB'* is automatically mapped to our *ParameterProfileDB* class, while *metavar='INT'* is mapped to *ParameterInt*. This approach allows tool developers to embed Galaxy-specific context directly into their command-line definitions in a non-intrusive way. Table 1 provides representative examples of this mapping, illustrating how specific *metavar* strings are translated into corresponding parameter classes and Galaxy-specific attributes.

| Metavar Attribute | Parameter Class | Galaxy Type | Format Attribute |
|---|---|---|---|
| CONTIGS_DB | ParameterContigsDB | data | anvio_contigs_db |

| PROFILE_DB | ParameterProfileDB | data | anvio_profile_db |
|---|---|---|---|
| PROFILE_DB_OUT | ParameterProfileDBOut | data | anvio_profile_db |
| PAN_DB | ParameterPanDB | data | anvio_pan_db |
| GENOMES_DB | ParameterGenomesDB | data | anvio_genomes_db |
| COLLECTION | ParameterCollection | data | anvio_collection |
| BIN | ParameterBin | data | anvio_bin |
| FASTA | ParameterFast | data | fasta |
| BAM | ParameterBam | data | bam |
| GENBANK | ParameterGenbank | data | genbank |
| TREE | ParameterTree | data | newick |
| TAXONOMY | ParameterTaxonomy | data | tabular |
| TABULAR | ParameterTabular | data | tabular |
| GFF | ParameterGFF | data | gff |
| VCF | ParameterVCF | data | vcf |
| FILE_PATH | ParameterFilePath | data | data |
| DIR_PATH | ParameterDirPath | data | directory |
| DIR_PATH_OUT | ParameterDirPathOut | data | directory |
| INT | ParameterInt | integer | N/A |
| FLOAT | ParameterFloat | float | N/A |
| STRING | ParameterString | text | N/A |

**Table 1:** Representative examples of *metavar* mapping to parameter classes and Galaxy XML attributes. The table illustrates the core convention of the framework, where a *metavar* string in a Python tool's *argparse* definition is used to determine the behavior and type of the corresponding parameter in the generated Galaxy tool. <u>Metavar Attribute</u>: the string provided to the *metavar* argument in the tool's *add_argument* function; <u>Parameter Class</u>: the internal Python class assigned by the framework to handle the specific logic for this parameter type; <u>Galaxy Type</u>: the value assigned to the *type* attribute of the *<param>* tag in the generated XML (e.g., *data*, *integer*, *boolean*). <u>Format Attribute</u>: the value assigned to the *format* attribute of the *<param>* tag, specifying the Galaxy datatype (e.g., *fasta*, *tabular*)

For arguments that lack a defined *metavar*, the classification engine falls back to a secondary mapping based on the argument's name (e.g., *--num-threads*) or its defined *action* (e.g., *action='store_true'*). This ensures that common, conventionally named arguments and simple boolean flags are still handled correctly even without explicit *metavar* declarations.

To manage the diverse requirements of different Galaxy parameters, we developed a hierarchy of Python classes inheriting from a single base *Parameter* class. Each specialized class encapsulates the logic needed to generate the correct XML for a specific parameter type and to construct its portion of the command-line call in the final wrapper. For instance, the *ParameterFILE_PATH* class handles standard data inputs and outputs, while more specialized classes like *ParameterContigsDB* and *ParameterProfileDB* contain additional logic to manage complex, composite datatypes. These classes can generate pre- and post-processing shell commands, such as creating output directories or linking input files, which are essential for handling unique file structures of tool's databases within the Galaxy environment. This object-oriented design makes the system extensible, as support for new or custom data types can be added simply by creating a new parameter class.

**Anatomy of a Parameter Class**

In order to illustrate the object-oriented design, we can examine the implementation of a specialized parameter class in detail. The *ParameterProfileDBOut* class is a prime example, as it is designed to handle the creation of an anvi'o profile database, a composite data type consisting of a SQLite database file and an associated auxiliary directory. This requires not only defining the output in the Galaxy interface but also injecting specific shell commands to correctly prepare the environment and collect the results. The key methods of this class are as follows:

- Inheritance and initialization: the class inherits from the base *Parameter* class. Its *__init__* method takes the parsed *argparse* object as input and sets key attributes, such as *galaxy_format='anvio_profile_db'* and a boolean flag *is_output=True*, which dictates its role in the XML generation;

- *to_xml_output()*: this method is responsible for generating the *<data>* tag within the *<outputs>* section of the Galaxy XML. It uses the stored attributes (name, format, label) from the *argparse* object to construct a well-formed XML element (e.g. *<data name="output_db" format="anvio_profile_db" />*). Because this is an output parameter, it does not generate a corresponding *<param>* tag in the *<inputs>* section;

- *to_cmd_line()*: this method generates the portion of the command-line call specific to this argument. For a composite output, it defines a predictable, temporary filename (e.g., *'anvio_profile.db'*) that the tool will write to within its isolated job directory. The generated snippet would be, for example, *-o 'anvio_profile.db'*. This ensures the output can be reliably found by subsequent post-processing steps;

- *get_pre_command()*: this is where the class's specialized knowledge is most apparent. It returns a string containing a shell command (e.g., *mkdir -p 'anvio_profile.d'*) that is injected into the *<command>* block before the main tool executable is called. This command creates the necessary auxiliary directory that the anvi'o program expects to find, preventing a runtime error;

- *get_post_command()*: systematically, this method returns a shell command that runs after the main tool has successfully executed. The command it generates (e.g., *cp -r 'anvio_profile.db'* '$output_db'*) is responsible for moving both the newly created SQLite database and its auxiliary directory into the final, empty dataset that Galaxy has prepared for the history item. This automated management of pre- and post-processing logic is what elevates the framework from a simple syntax translator to a context-aware wrapper generator.

This collection of methods, encapsulating within a single class, ensures that every aspect of handling this complex output, from its declaration in the user interface to its final placement in the user's history, is managed automatically and correctly. It is this object-oriented design that allows the framework to be easily extended to support new, custom data types by simply defining a new class with the appropriate logic, without altering the core wrapper generation engine.

**Generation of Galaxy Tool XML**

The final stage of the framework translates the classified parameter objects into complex, well-formed Galaxy tool XML files. This process is driven by the Jinja2 templating engine, which populates a master *TOOL_TEMPLATE* with the specific details extracted from the target tool. This template defines the static structure of a Galaxy wrapper, including the root *<tool>* tag and standard sections for requirements, citations, and help.

The dynamic sections of the wrapper (i.e., *<input>*, *<output>*, and *<command>*) are constructed programmatically. The object-oriented design of the parameter classes is central to this process. Each specialized parameter object contains methods (*to_xml_param()* and *to_xml_output()*) that return the precise XML string for its corresponding input or output element. The framework iterates through the list of classified parameter objects and concatenates these strings to build the complete *<input>* and *<output>* blocks, ensuring that all necessary attributes (e.g., *type*, *format*, *label*) are correctly defined.

Similarly, the *<command>* block is generated by calling a *to_cmd_line()* method on each parameter object. This method returns the argument's command-line flag followed by the appropriate Galaxy-specific variable (e.g., *--input-file '${my_input}'*). This approach seamlessly handles the logic for optional parameters by wrapping the command-line snippets in a Cheetah [31] *#if…#end if* block, which is the templating language used within the Galaxy *<command>* section. Finally, static information such as the tool's help text, which is extracted from the *argparse description* and *epilog*, is formatted and inserted into the *<help>* section, resulting in a complete, human-readable, and machine-executable Galaxy tool wrapper.

The complete software architecture, from script introspection to final XML generation, is summarized in Figure 1 below.

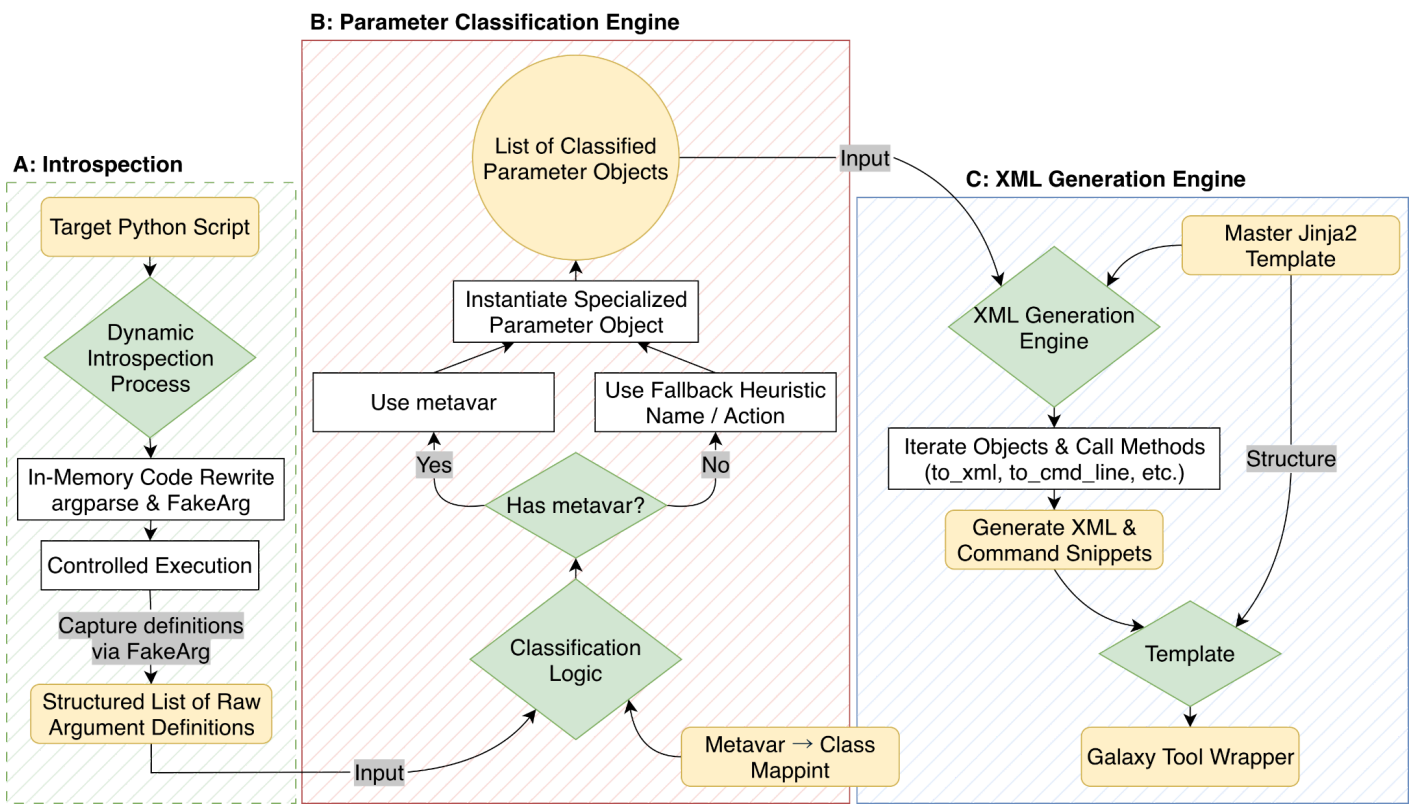

**Figure 1:** Architectural overview of the automated Galaxy tool wrapper generation framework. The figure illustrates the end-to-end pipeline for converting a Python command-line tool into a valid Galaxy XML wrapper. **Panel A** – Introspection engine: the process begins with a target Python script. The framework reads the script, replaces the standard *argparse* library with a custom *FakeArg* class in-memory, and executes the modified code to capture all argument definitions without running the tool's main logic; **Panel B** – Parameter classification engine: the captured arguments are fed into a classification engine. Each argument is mapped to a specialized parameter class using a multi-tiered system that first check for a defined *metavar*, then falls back to the argument's name or action; **Panel C** – XML generation engine: the resulting list of classified parameter objects is used by a Jinja2 templating engine to populate a master XML template. The engine programmatically constructs the *<input>*, *<output>*, and *<command>* sections, yielding a complete and functional Galaxy tool wrapper ready for deployment.

## Automated Management of Software Dependencies

A critical and often laborious aspect of creating robust Galaxy tools is the precise definition of all software dependencies. A fundamental prerequisite for any tool integration into Galaxy is the availability of all its software dependencies through the Conda package manager. Before attempting to use our framework, developers must first ensure that every dependency required by their tool is available as a Conda package. If a dependency is not available, it must first be packaged and contributed to a public Conda channel (such as Bioconda), a process that is outside the scope of this work.

Once all dependencies are confirmed to be available, our framework can automate the generation of the *<requirements>* section of the Galaxy tool wrapper. The process is initiated by scanning the root directory of the target tool's project for standard Python dependency manifest files, such as *requirements.txt* or a Conda *environment.yml*. The framework prioritizes Conda environment files when available, as they provide explicit versioning and channel information that maps directly to Galaxy's use of the Conda package manager for dependency resolution.

Upon locating a dependency file, the framework parses it to extract a list of all required packages and their specified versions. Each package is then translated into a *<requirement>* tag in the final XML. For example, a line like *pandas=1.4.2* in an *environment.yml* file would be converted to *<requirement type="package" version="1.4.2">pandas</requirements>*. This automated process eliminates the risk of human error associated with manually transcribing dozens of dependencies and ensures that the generated Galaxy tool is a self-contained, fully reproducible scientific appliance.

**A Roadmap for Developer Adoption**

In order to facilitate the adoption of this framework by the broader scientific software community, we provide a straightforward roadmap for developers who wish to make their Python-based tools compatible with our automated wrapping pipeline. The process is designed to be minimally intrusive and requires only minor additions to a tool's existing source code.

Step 1: Prerequisites – The primary requirement is that the target tool is written in Python and uses the standard *argparse* library for command-line argument parsing. No other dependencies are required for the tool itself.

Step 2: Adopting the *metavar* convention – The core of the integration process is the adoption of the *metavar* convention for input and output arguments that correspond to data files. For each data-handling argument in the *argparse* definition, the developer should add a *metavar* string that semantically describes the data type. The comprehensive list of recognized *metavar* attributes and their corresponding Galaxy types is provided in Table 1.

It is important to first ensure that a parameter class corresponding to the desired *metavar* is implemented within our framework. If a new, unsupported data type is needed, the framework's object-oriented design makes it straightforward to add one. Developers can implement their own custom class by following the structure outlined in the "Anatomy of a Parameter Class" section presented above, ensuring that even novel data types can be seamlessly integrated.

For example, consider a simple argument for a FASTA file input. A typical definition might look like the following one.

> *Before:*
> *parser.add_argument('--input-seqs', required=True, help='Path to the input sequences.')*

To make this argument compatible with our framework, the developer should simply add the corresponding *metavar* string.

> *After:*
> *parser.add_argument('--input-seqs', required=True, help='Path to the input sequences.',*
> ***metavar='FASTA')***

This simple addition provides the framework with the necessary context to correctly identify the parameter as a FASTA data input in Galaxy;

Step 3: Running the Wrapper Generator – Once the *metavar* attributes are in place, the developer can run the generator script, providing the path to their tool as input. The script will execute the introspection and classification engine as described previously;

Step 4: Deployment – The output of the script is a standard, well-formed Galaxy tool XML file. This wrapper is ready for deployment and can be integrated into a local Galaxy instance or uploaded to the public Galaxy ToolShed for community use. As the automated generation of test cases is not yet supported (see Current Limitations section for additional details), developers are strongly encouraged to manually test the wrapper's functionality before public release.

## RESULTS

To validate the efficacy and scalability of our automated framework, we applied it to the *anvi'o* software suite, a large and complex bioinformatics platform. The following sections detail the quantitative and qualitative outcomes of this application, demonstrating the framework's ability to successfully generate high-quality, functional Galaxy tool wrappers from command-line Python scripts with no manual intervention.

### Wrapper Generation for the *anvi'o* Suite

The automated wrapping framework was executed on the *anvi'o* source code (v7), targeting all executable Python scripts within the primary *bin* and secondary *sandbox* directories. This collection of scripts represents the complete user-facing functionality of the *anvi'o* platform. Out of a total of 176 Python scripts in the *anvi'o* package, the framework successfully parsed the *argparse* definitions and generated 148 corresponding Galaxy tool wrappers. The small number of scripts that were not converted were intentionally excluded via a predefined skip-list (e.g., *anvi-upgrade*, a script for managing *anvi'o* installations) as not suitable for the Galaxy environment.

### Handling of Complex *anvi'o* Data Structures and Interactive Tools

A key challenge presented by the *anvi'o* suite is its reliance on complex, composite data structures, where a single dataset consists of a primary database file and an associated directory of auxiliary files. Our framework successfully addresses this challenge through its specialized classes. When a parameter is identified as a composite type (e.g., *ParameterProfileDB* or *ParameterContigsDB*), the class automatically generates the necessary shell commands to manage these structures within Galaxy's isolated job environment. For an output database, the framework injects a *mkdir* command into the *<command>* block to create the required extra files directory. For an input database that will be modified, the framework generates a *cp -R* command to safely copy the input composite dataset to a new output dataset, ensuring the original history item remains unchanged. This automated handling of pre- and post-processing steps is critical for the correct functioning of many *anvi'o* tools.

In addition to standard command-line tools, the framework proved capable of wrapping *anvi'o*'s interactive visualization tools. During the introspection phase, the system checks for the presence of a *__provides__* variable in the script's global namespace containing the string *interactive*. When this condition is met, the framework assigns the tool an *interactive* type. This classification triggers the XML generator to populate the wrapper with an *<entry_points>* section, which is required to make the tool compatible with Galaxy's Interactive Environment feature. This result demonstrates that the framework's introspection capabilities extend beyond *argparse* definitions, allowing it to capture additional metadata from the script to handle advanced Galaxy features automatically.

**Validation of Generated Wrappers**

To ensure the generated wrappers are not only functional but also adhere to community standards, the entire set of 148 XML files was validated using *planemo lint*, the standard command-line utility for Galaxy tool quality control. The automated validation confirmed that the wrappers were well-formed and structurally correct. The process identified one warning, such as the absence of the example test cases, which are outside the scope of automated generation from interface parsing alone.

**Case Study 1: standard reporting tool**

To qualitatively assess the framework's output, we first examine the wrapper generated for *anvi-summarize*, a key *anvi'o* program used to create summary reports from profile databases. This tool serves as a representative example of a standard, non-interactive program with multiple input data types and optional flags. The framework correctly translates the *argparse* definitions from the *anvi-summarize* Python script into corresponding Galaxy XML elements. For instance, the definition for the primary input database in the Python source code is:

```
parser.add_argument('-p', '--profile-db', dest='profile_db', required=True, help='A profile database', metavar='PROFILE_DB')
```

Our framework parses this line and, using the *metavar* to *Parameter* class mapping, generates the following XML element for the Galaxy user interface:

```
<param name="profile_db" type="data" format="anvio_profile_db" label="Profile Db" argument="--profile-db" help="A profile database" />
```

This demonstrates the accurate translation of the argument's name, help text, and command-line flag. More importantly, it shows the successful mapping of the *PROFILE_DB metavar* to the specific *anvio_profile_db* Galaxy datatype, ensuring proper data handling.

The tool also includes optional boolean flags, such as *--init-gene-coverages*, which instructs the program to perform an additional calculation. The *argparse* definition for this flag is:

*parser.add_argument('--init-gene-coverages', dest='init_gene_coverages', action='store_true', help='Initialize gene coverages table.')*

The framework recognizes the *action='store_true'* and generates a boolean parameter in the Galaxy interface. Crucially, it also constructs the appropriate Cheetah templating logic in the *<command>* block to ensure the flag is only added to the command line if the user selects "Yes" in the interface, as shown in Code Box 1 below.

```
<command>
    <![CDATA[
        anvi-summarize --profile-db '$profile_db'
        #if $init_gene_coverages:
            --init-gene-coverages
        #end if
        --output-dir 'output' &&
        cp -r output/* '$output_file'
    ]]>
</command>
```

**Code Box 1:** Automated generation of conditional command-line logic. This XML snippet illustrates the *<command>* block generated for *anvi-summarize*, demonstrating how the framework wraps the optional *--init-gene-coverages* flag within a Cheetah *#if* block. This logic ensures the argument is dynamically added to the command line only when the user enables the corresponding boolean parameter in the Galaxy interface.

This example shows how the system correctly assembles the base command, conditionally appends optional parameters, and handles the collection of output files, creating a robust and functional command-line call based on user input from the Galaxy interface.

The resulting user interface for *anvi-summarize* is displayed in Figure 2, confirming that the framework correctly renders the various input types and optional flags defined in the source code.

🔧 **anvi-summarize** ☆ ▾ ▶ Run Tool

Summarizer for anvi'o pan or profile db's. Essentially, this program takes a collection id along with either a profile database and a contigs database or a pan database and a genomes storage and generates a static HTML output for what is described in a given collection. The output directory will contain almost everything any downstream analysis may need, and can be displayed using a browser without the need for an anvi'o installation. For this reason alone, reporting summary outputs as supplementary data with publications is a great idea for transparency and reproducibility
(Galaxy Version 7.1)

**Tool Parameters**

❗ Please provide a value for this option.
**Pan Or Profile Db** * required

| 📄 | 📑 | 📁 | ... | No anvio_profile_db or anvio_pan_db datasets available |

accepted formats ▾

**(i) Specialized data inputs:**
"Pan Or Profile Db" restricts inputs to specific formats (anvio_profile_db, anvio_pan_db) using the framework's classification engine

Anvi'o pan or profile database (and even genes database in appropriate contexts). (--pan-or-profile-db)

**Contigs Db** - optional

| 📄 | 📑 | 📁 | ... | Nothing selected | ▾ |

accepted formats ▾

Anvi'o contigs database generated by 'anvi-gen-contigs-database' (--contigs-db)

**(ii) Boolean parameters:**
"Initial Gene Coverages" renders as a toggle switch, automatically derived from *action="store_true"* arguments

**Genomes Storage** - optional

| 📄 | 📑 | 📁 | ... | Nothing selected | ▾ |

accepted formats ▾

Anvi'o genomes storage file ( ...nes-storage)

**(iii) Integrated documentation:**
Help text extracted from the Python script is displayed directly below each parameter alongside the original command-line flag

**Init Gene Coverages**

⚪ No

Initialize gene coverage and detection data. This is a very computationally expensive step, but it is necessary when you need gene level coverage data. The reason this is very computationally expensive is because anvi'o computes gene coverages by going back to actual coverage values of each gene to average them, instead of using contig average coverage values, for extreme accuracy. (--init-gene-coverages)

**Reformat Contig Names**

⚪ No

Reformat contig names while generating the summary output so they look fancy. With this flag, anvi'o will replace the original names of contigs to those that include the bin name as a prefix in resulting summary output files per bin. Use this flag carefully as it may influence your downstream analyses due to the fact that your original contig names in your input FASTA file for the contigs database will not be in the summary output. Although, anvi'o will report a conversion map per bin so you can recover the original contig name if you have to. (--reformat-contig-names)

**Report Aa Seqs For Gene Calls**

⚪ No

You can use this flag if you would like amino acid AND dna sequences for your gene calls in the genes output file. By default, only dna sequences are reported. (--report-aa-seqs-for-gene-calls)

**Report Dna Sequences**

⚪ No

By default, this program reports amino acid sequences. Use this flag to report DNA sequences instead. Also note, since gene clusters are aligned via amino acid sequences, using this flag removes alignment information manifesting in the form of gap characters ('-' characters) that would be present if amino acid sequences were reported. Read the warnings during runtime for more detailed information. (--report-DNA-sequences)

**Collection Name** - optional

| | ⌄ |

Collection name. (--collection-name)

**List Collections**

⚪ No

Show available collections and exit. (--list-collections)

**Cog Data Dir** - optional

| 📄 | 📑 | 📁 | ... | Nothing selected | ⌄ |

accepted formats ▾

The directory path for your COG setup. Anvi'o will try to use the default path if you do not specify anything. (--cog-data-dir)

**Quick Summary**

⚪ No

When declared the summary output will be generated as quickly as possible, with minimum amount of essential information about bins. (--quick-summary)

**Figure 2:** The automatically generated Galaxy tool interface for *anvi-summarize*. This demonstrates the framework's capability to translate complex *argparse* definitions into a native Galaxy GUI without manual intervention. Key elements include: (i) specialized data inputs, such as the "Pan Or Profile Db" parameter, which uses the framework's classification engine to restrict inputs to specific formats (*anvio_profile_db* or *anvio_pan_db*) based on the *PROFILE_DB* metavar in the source code; (ii) boolean parameters, such as "Init Gene Coverages", which are rendered as toggle switches derived automatically from arguments defined with *action='store_true'*; and (iii) integrated documentation, where the help text extracted from the Python script is displayed directly below each parameter alongside the original command-line flag (e.g., *--init-gene-coverages*) to ensure transparency for the user.

## Case Study 2: handling complex data structures

A more significant challenge presented by the anvi'o suite is its reliance on complex, composite data structures, where a single dataset consists of a primary database file and an associated directory of auxiliary files. The *anvi-profile* program, which generates a profile database from sequence alignments, is a prime example. This process requires not only parsing input files but also creating a new composite data structure as output.

Our framework successfully addresses this challenge through its specialized parameter classes. The *argparse* definition for the output profile database is:

>*parser.add_argument('-o', '--output-db', dest='output_db', help='Path to the output profile database.', metavar='PROFILE_DB_OUT', required=True)*

The *PROFILE_DB_OUT metavar* maps to the *ParameterProfileDBOut* class. This specialized class understands that a profile database requires an auxiliary directory. Consequently, it injects the necessary pre-processing shell commands into the *<command>* block. The generated XML automatically includes a *mkdir* command to create the required directory before the main tool is executed and a *cp* command to collect the contents into the final Galaxy history item.

The generated *<command>* block in Code Box 2 illustrates this process.

```
<command>
    <![CDATA[
    #set $output_db_path = "anvio_profile.db"
    mkdir -p '${output_db_path}.d' &&
    anvi-profile --input-file '$input_file'
                 --contigs-db '$contigs_db'
                 -o '$output_db_path'
                 --sample-name '$sample_name' &&
    cp -r '${output_db_path}'* '$output_db'
    ]]>
</command>
```

**Code Box 2:** Automated management of composite data structures. The generated XML for *anvi-profile* illustrates the injection of pre- and post-processing shell commands. The framework automatically inserts a *mkdir* command to initialize the required auxiliary directory structure before execution and a *cp* command to correctly capture the composite output (database and directory) into the final Galaxy history item.

This automated handling of pre- and post-processing steps is critical for the correct functioning of many *anvi'o* tools and demonstrates the framework's ability to manage complex file structures without manual intervention. It elevates the system from a simple argument-to-XML translator to a context-aware wrapper generator.

Figure 3 illustrates the resulting interface.

**Figure 3:** The automatically generated interface for the *anvi-profile* tool. This demonstrates the framework's capacity to handle extensive and complex command-line interfaces without manual wrapping. Presented in two columns for readability, the interface showcases several key features: (i) strongly-typed data inputs (left panel), where the *Input File* is correctly restricted to BAM format and *Contigs Db* to the specialized *anvio_contigs_db* format; (ii) diverse parameter mapping (right panel), illustrating the automated rendering of integers (e.g., "Min Contig Length"), dropdown selections (e.g., "Distance"), and boolean toggles (e.g., "Cluster Contigs"); and (iii) abstracted complexity, where the "Run Tool" button triggers the hidden *ParameterProfileDBOut* logic that executes the necessary *mkdir* and *cp* commands to successfully generate and capture the tool's composite output structure (a database file plus an auxiliary directory) within the Galaxy history.

**Case Study 3: supporting interactive visualization tools**

In addition to standard command-line tools, our framework proved capable of wrapping *anvi'o* interactive visualization tools. These tools, which launch a web server to provide a rich, user-driven interface, represent a distinct class of application that requires special handling within Galaxy. The *anvi-interactive* tool is the flagship visualization component of the suite.

During the introspection phase, the system implements a convention to detect these tools: it checks for the presence of a *__provides__* variable in the script's global namespace containing the string *interactive*. When this condition is met, the framework assigns the tool an *interactive* type. This classification triggers the XML generator to populate the wrapper with an *<entry_points>* section, which is required to make the tool compatible with Galaxy's Interactive Environment feature.

The generated XML for anvi-interactive includes the section in Code Box 3 below.

```
<entry_points>
    <entry_point name="Anvi'o Interactive" port="8080">
        <url>interactives/anvio</url>
    </entry_point>
</entry_points>
```

**Code Box 3:** Automated generation of interactive entry points. This XML snippet shows the *<entry_points>* section generated for *anvi-interactive*. By detecting the tool's interactive nature during introspection, the framework automatically defines the necessary port and URL configuration, allowing Galaxy to provision a web server and embed the visualization directly within the user interface.

This block instructs Galaxy on how to connect to the running *anvi'o* web server, enabling the visualization to be displayed within the user's workspace. This result demonstrates that the framework's introspection capabilities extend beyond *argparse* definitions, allowing it to capture additional metadata from the script to handle advanced Galaxy features automatically. This capability, summarized in Figure 4, is crucial for integrating modern, complex bioinformatics platforms that blend command-line processing with interactive data exploration.

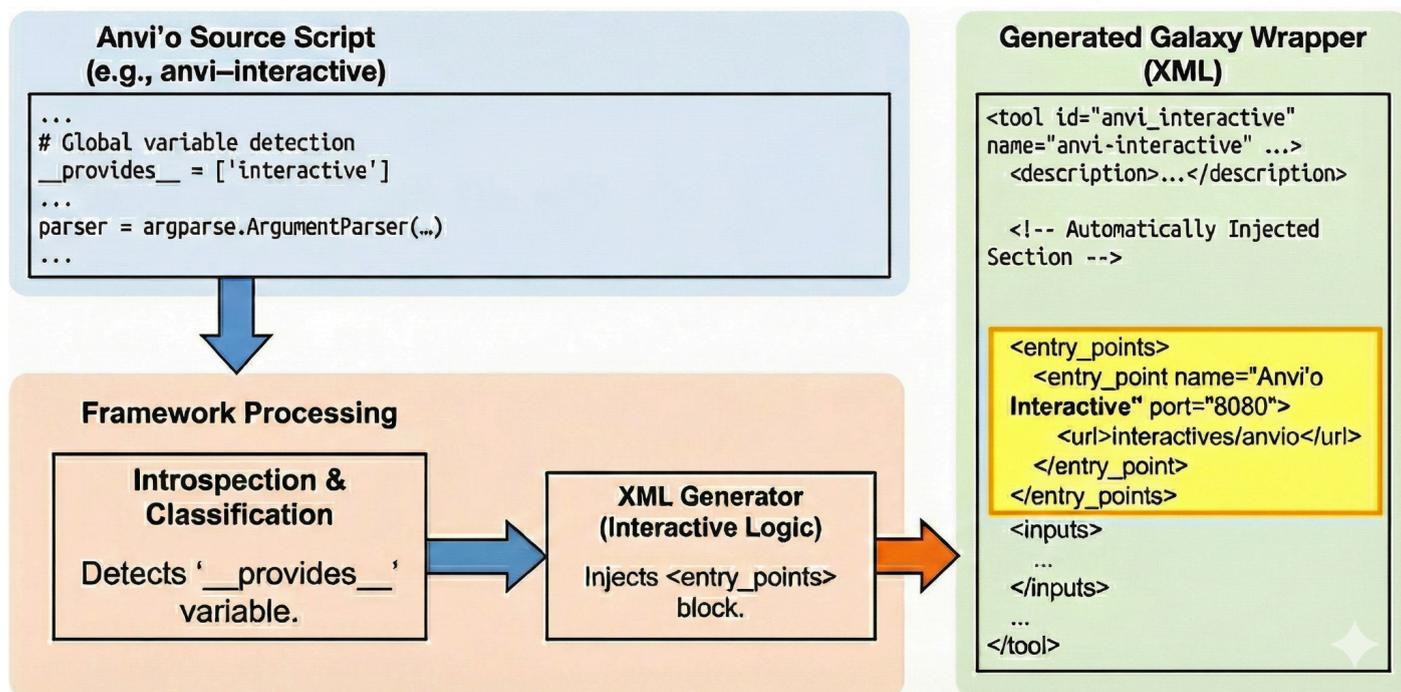

**Figure 4:** Schematic representation of the automated generation of Galaxy wrappers for interactive *anvi'o* tools. This diagram illustrates the workflow through which the framework automatically detects and wraps *anvi'o* tools that require interactive web-based environments, removing the need for manual XML configuration. Anvi'o Source Script (blue panel) – the process begins with an existing Python script for an *anvi'o* tool. The code snippet highlights a critical metadata convention used by the framework: the definition of a global variable "*__provides__*" containing the string "*interactive*". This flag serves as the indicator that the tool launches a local web server for interactive visualization; Framework Processing (orange panel) – this details the automated actions taken by the framework upon receiving the source script. First, the Introspection & Classification module parses the script's global namespace and detects the specific "*__provides__*" variable defined in the blue panel. Based on this detection, the tool is classified as "interactive". This classification then triggers the specialized XML Generator (Interactive Logic), which is tasked with injecting the specific XML structure required to support Galaxy Interactive Environments; Generated Galaxy Wrapper (green panel) – this shows the resulting Galaxy tool wrapper command file. The yellow-highlighted box emphasizes the automatically injected section created by the interactive logic in the middle panel. This injected *<entry_points>* block defines the necessary configurations, specifically enabling access on *port="8080"* and setting the relative URL path to *<url>interactives/anvio</url>*, allowing Galaxy to dynamically provision a web server and embed the tool's interactive interface within the Galaxy user experience.

## Case Study 4: reproducing a full metagenomic workflow

The ultimate test of the framework is not whether individual tools function in isolation, but whether the entire integrated suite can be used to perform a complex, multi-step scientific workflow. To demonstrate this capability, we reproduced the canonical anvi'o metagenomic workflow, from raw contigs to a final, summarized profile database, entirely within the Galaxy environment. This workflow is a cornerstone of *anvi'o*-based analyses and serves as a real-world test of the tool suite's interoperability.

The analysis was performed as a standard Galaxy workflow, chaining together the outputs of one tool as the inputs for the next, with each step corresponding to a major stage in the official tutorial available at https://anvio.org/help/main/workflows/metagenomics/:

1. <u>Generating the contigs database</u>: the workflow began with a FASTA file of assembled contigs from a metagenomic sample [32]. The *anvi-gen-contigs-database* tool was used to create the foundational contigs database. The crucial first step annotates the contigs by identifying open reading frames [33], searching for common single-copy core genes [34], and calculating tetranucleotide frequencies, preparing the data for all downstream analyses;

2. <u>Profiling individual samples</u>: next, several BAM files [35], representing the mapping of short reads from a specific sample back to the assembled contigs, was used as input. The *anvi-profile* tool was executed on each BAM file to generate an individual *anvi'o* profile database. This step calculates the coverage and detection statistics for each contig across each sample, which is essential for differential abundance analysis [36] and binning;

3. <u>Merging profiles for co-analysis</u>: the individual profile databases were then combined into a single, merged profile database using the *anvi-merge* tool. This step is critical for creating a unified view of all samples, enabling the subsequent algorithms to leverage co-variation patterns across the entire dataset;

4. <u>Metagenomic binning</u>: with the merged profile, we proceeded to the binning stage to identify putative genomes. The *anvi-cluster-contigs* tool was used, employing the CONCOCT algorithm [37] to group contigs into bins based on their sequence composition and differential coverage patterns across the samples. The output was a collection of bins, representing candidate MAGs [38,39];

5. <u>Summarization and interactive refinement</u>: finally, the *anvi-summarize* tool was run on the merged profile and binning results to generate a comprehensive tabular summary of the MAGs, including their taxonomy [40], completeness, and redundancy estimates [41].

The successful execution of this entire workflow within Galaxy (Figure 5) provides the most compelling validation of our framework. It demonstrates that the automatically generated wrappers correctly handle not only the parameters of each tool but also the complex, composite data types that are passed between them. The analysis, mirroring an official, expert-curated tutorial, was completed without writing a single line of code, and the resulting Galaxy history serves as a complete, transparent, and reproducible record of the entire metagenomic analysis.

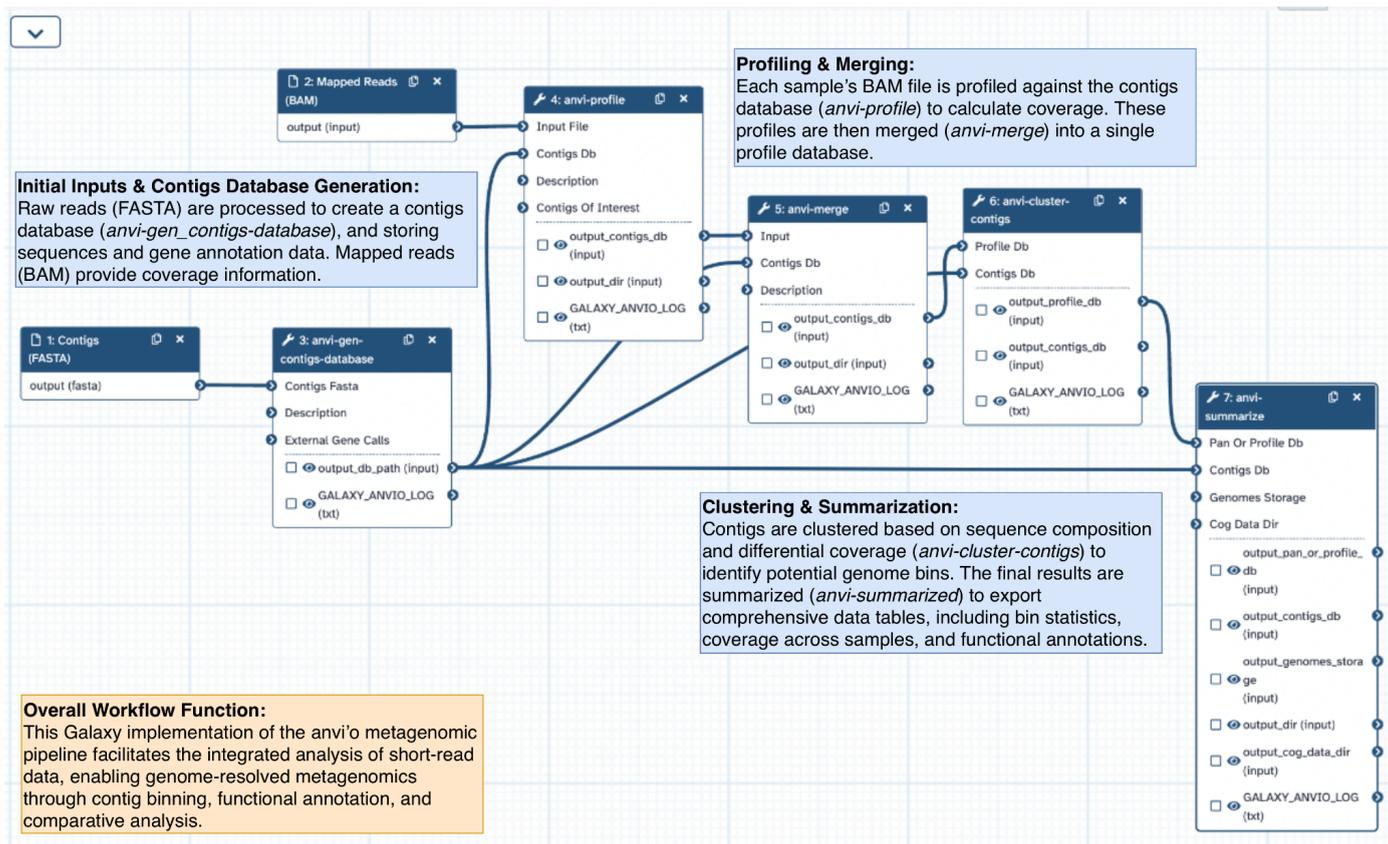

**Figure 5:** The complete *anvi'o* metagenomic workflow reproduced in Galaxy. This annotated workflow diagram illustrates the successful execution of the multi-step metagenomic analysis described in Case Study 4. The pipeline is divided into three logical phases: (i) <u>Contigs Database Generation</u> (left), where raw FASTA contigs are processed by *anvi-gen-contigs-database* to generate the foundational *anvio_contigs_db* containing gene calls and k-mer frequencies; (ii) <u>Profiling & Merging</u> (center), where sample-specific BAM files are processed by *anvi-profile* and subsequently combined via *anvi-merge* to create a unified profile database; and (iii) <u>Clustering & Summarization</u> (right), where the merged profile is binned using *anvi-cluster-contigs* and finally exported as a static HTML summary using *anvi-summarize*. The visual connectivity, specifically the explicit wiring of the Contigs Db output from step 3 to every subsequent tool (steps 5, 5, 6, and 7), confirms that the automated wrappers correctly expose and manage complex, composite data types across the entire analysis chain.

## DISCUSSION AND CONCLUSIONS

In this work, we have presented a novel, automated framework for generating Galaxy tool wrappers directly from the source code of Python-based tools. Our approach directly confronts the bottleneck of integrating tools, a long-lasting challenge that affected the rapid deployment of new analytical methods onto the Galaxy platform. By programmatically introspecting a tool's *argparse* definition, our system effectively translates a command-line interface into a fully-featured graphical user interface, drastically reducing the manual effort, time, and specialized knowledge required for tool wrapping.

The primary significance of this framework lies in its potential to accelerate the scientific discovery process. The ability to automatically generate functional wrappers means that new and updated tools can be made available to the broad community of Galaxy users in a fraction of the time previously required. This not only empowers scientists and other domain experts with access to cutting-edge tools but also enhances research by promoting the use of standardized, reproducible workflows. By lowering the barrier to integration, our method encourages developers to make their tools available on the Galaxy platform, directly supporting the FAIR data principles.

The core strength of our methodology is its generalizability, which stems from its reliance on ubiquitous Python libraries [42] and a simple, non-intrusive convention. By leveraging *argparse*, a standard component of the Python ecosystem, the framework is immediately applicable to a vast number of existing scientific tools without requiring developers to adopt a new, complex API or rewrite their argument parsing logic. The use of the optional *metavar* attribute as the primary mechanism for semantic type declaration is a key design choice, as it allows developers to embed Galaxy-specific context directly into their code in a way that is both human-readable and machine-parsable, without affecting the tool's standard command-line functionality.

Beyond its immediate utility, our framework has broader implications for best practices in scientific software development. The *metavar* convention effectively establishes a standard for creating "workflow-ready" tools. By encouraging developers to embed semantic and structural metadata directly into the command-line interface, this approach promotes the creation of tools that are inherently more findable, accessible, and interoperable. Adopting such conventions from the outset improves a tool's documentation and usability for all users, not just those on the Galaxy platform, and fosters a culture where interoperability is considered during development, not as an afterthought. This shift aligns tool development more closely with the FAIR principles, benefiting the entire computational ecosystem.

## Current Limitations

It is also important to acknowledge the limitations of the current framework. Its applicability is presently confined to tools written in Python, leaving a significant number of tools written in other languages outside its scope. Furthermore, the precision of the automated wrapping is highest when developers adhere to the *metavar* convention. While the system includes fallback mechanisms that classify parameters based on their names or actions, these are inherently less precise and may not correctly interpret ambiguously named arguments. The framework may also struggle with highly customized or non-standard implementations of *argparse* that deviate significantly from common usage patterns. Finally, the creation of automated test cases, a crucial component for validation within the Galaxy ecosystem, is not addressed and still requires manual effort.

## Future Directions

These limitations discussed above highlight several promising avenues for future work. The immediate next step is to extend the framework's capabilities to other programming languages and their respective argument-parsing libraries. A more ambitious direction would be to develop a more sophisticated type-inference engine that could analyze not just the argument definition but also its usage within the code, thereby reducing the reliance on the *metavar* convention. We also envision a semi-automated approach to test generation [43], where the framework could create placeholder test definitions, further reducing the manual burden on developers.

In conclusion, the framework presented here offers a practical and scalable solution to the critical challenge of tool integration into the Galaxy platform. By automating the conversion of command-line tools into accessible graphical applications, our work helps to close the gap between tool development and its application in scientific research, promoting a more agile, accessible, and reproducible computational ecosystem.

## ADDITIONAL INFORMATION

### Availability

The source code for the automated tool wrapping framework is open-source and publicly available on GitHub under the MIT License at https://github.com/BlankenbergLab/tool-generator-anvio. The repository contains all scripts required to reproduce the work described in this paper, along with detailed documentation.

### Author Contributions

DB conceived the research; FB and DB designed the methodology; FC and DB implemented the pipeline and performed the analysis; FC and JJ tested and validated the pipeline; FC, JJ, and DB wrote the manuscript and agreed with its final version.

### Conflict of Interests

DB has a significant financial interest in GalaxyWorks, a company that may have a commercial interest in the results of this research and technology. This potential conflict of interest has been reviewed and is managed by the Cleveland Clinic Foundation.

FC and JJ have no conflicts to disclose.

### Funding

This work was supported by the Wellcome Trust [313498/Z/24/Z].

### Acknowledgments

The authors acknowledge the use of a Large Language Model (Google's Gemini 3 Pro) for the sole purpose of enhancing the clarity and readability of this manuscript. The scientific ideas, methodology, and implementation presented herein are entirely the product of the authors' experience and expertise.